\documentclass[epj]{svjour}

\usepackage{graphicx}
\usepackage{fancyhdr}
\usepackage{amssymb}
\def\makeheadbox{{
 }}
\def\mytitle{My title} 
\def\myauthors{My name}  
\def\mytype{My type of session}
\def\mysession{My session}

\def\mytitle{$H^\pm W^\mp$ production in the MSSM at the LHC}
\def\myauthors{D.~Eriksson, S.~Hesselbach, J.~Rathsman}
\def\mytype{Contributed Talk}
\def\mysession{Colliders - Higgs Phenomenology}

\begin{document}
\title{Associated charged Higgs and \boldmath$W$ boson production in the MSSM
  at the LHC}
\author{David~Eriksson\inst{1}
 \and
 Stefan~Hesselbach\inst{2}
\thanks{\emph{Email:} s.hesselbach@phys.soton.ac.uk}%
 \and
 Johan~Rathsman\inst{1}
}                     
%
%
\institute{High Energy Physics, Uppsala University,
Box 535, S-75121 Uppsala, Sweden
\and
School of Physics \& Astronomy, University of Southampton,
Highfield, Southampton SO17~1BJ, UK}
%
\date{}
\abstract{
We investigate the associated production of charged Higgs bosons
($H^\pm$) and $W$ bosons at the CERN Large Hadron Collider, 
using the leptonic decay $H^+ \to \tau^+ \nu_\tau$ and hadronic $W$ decay,
within different scenarios of the Minimal Supersymmetric Standard Model (MSSM)
with both real and complex parameters.
Performing a parton level study we show how the irreducible Standard Model
background from $W+2$ jets can be controlled by applying appropriate cuts.
In the standard $m_h^\mathrm{max}$  scenario we find a viable signal
for large $\tan\beta$ and intermediate $H^\pm$ masses ($\sim m_t$).
In MSSM scenarios with large mass-splittings among the heavy Higgs bosons 
the cross-section can be resonantly enhanced by factors up to one
hundred, with a strong dependence on the CP-violating phases. 
} 
\maketitle
\section{Introduction}

One of the main tasks at the upcoming
experiments at the CERN Large Hadron Collider (LHC) is the
determination of the mechanisms of electroweak symmetry breaking and of
the generation of elementary particle masses.
In the Minimal Supersymmetric Standard Model (MSSM),
which is a two Higgs Doublet Model (2HDM) of type II, the
Higgs sector consists of three neutral and one charged Higgs bosons
after electroweak symmetry breaking.
The charged Higgs boson ($H^\pm$) is of special interest since its discovery
would constitute an indisputable proof of physics beyond the Standard
Model (SM).

\sloppypar
The main production mode of charged Higgs bosons at hadron colliders is in
association with top quarks through the $gb \to H^-t$
and $gg \to H^-t\bar{b}$
processes~\cite{Barnett:1987jw,Bawa:1989pc,Borzumati:1999th,Miller:1999bm,Alwall:2004xw,Mohn:2007fd,Hesselbach:2007jj}
with the former one being dominant for heavy charged Higgs bosons
$m_{H^\pm} \gtrsim m_t$ and the latter one for light ones
$m_{H^\pm} \lesssim m_t - m_b$.
A complementary production mode of charged Higgs bosons is in
association with
$W$-bosons~\cite{Dicus:1989vf,BarrientosBendezu:1998gd,Moretti:1998xq,Brein:2000cv,Asakawa:2005nx,Eriksson:2006yt,Gao:2007wz}.
Although the production
cross-section~\cite{Dicus:1989vf,BarrientosBendezu:1998gd}
is large, an earlier study~\cite{Moretti:1998xq}
using the hadronic charged Higgs decay,
$H^+ \to t \bar{b}$, came to the conclusion that the signal is overwhelmed by
the $t \bar t$ background.

Here, we report results of our study~\cite{Eriksson:2006yt} of the
prospects of using instead
the $H^+ \to \tau^+ \nu_\tau$ decay together with $W \to 2 \textrm{ jets}$
in the MSSM with both real and complex parameters.
In the following we first outline the simulation of $H^\pm W^\mp$ production
and decay at the LHC and then give our results 
for the standard maximal mixing ($m_h^\mathrm{max}$)
scenario as well as for scenarios with resonantly
enhanced cross section.
For more details we refer to~\cite{Eriksson:2006yt}.

\section{Simulation and signal selection}

\begin{table*}
\centering
\caption{The effect of the different cuts on the integrated cross-section for
  background  ($\sigma_{\rm b}$) and signal  ($\sigma_{\rm s}$) in the
  $m_h^\mathrm{max}$ scenario
  with $m_{H^\pm}=175$ and 400 GeV for $\tan\beta=50$
  as well as the number of signal events $S$ and the significance $S/\sqrt{B}$
  assuming an integrated luminosity of
  300 fb$^{-1}$ and a $\tau$ detection efficiency of 30\% .
}\label{cuts}
\begin{tabular}{c@{\hspace{10mm}}c@{\hspace{10mm}}ccc@{\hspace{10mm}}ccc}
  \hline \hline
 & & \multicolumn{3}{c@{\hspace{10mm}}}{$m_{H^\pm}=175$ GeV} &
 \multicolumn{3}{c}{$m_{H^\pm}=400$ GeV}  \\
Cut [all in GeV] & $\sigma_{\rm b}$ (fb) & $\sigma_{\rm s}$ (fb) &
$S$ & $S/\sqrt{B}$& $\sigma_{\rm s}$ (fb) & $S$ & $S/\sqrt{B}$ \\ \hline
Basic cuts &
 560000 & 55 & 4900  & 0.7 & 3.3 & 300 & 0.04 \\
$p_{\perp\tau_\mathrm{jet}}>50$ , $\mbox{$\!\not \!p_{\perp}$}>50$ &
 22000 & 25 & 2200 &  1.6 & 2.7 & 240 & 0.2 \\
$70 <m_{jj}<90$  &
 1700 & 21 & 1900 &  5 & 2.2 & 200 & 0.5 \\
$m_\perp>100$  &
 77 & 15 & 1400 &  16 & 2.1 & 190 & 2.3 \\
$p_{\perp hj}>50$  , $p_{\perp sj}>25$  &
 28 & 9.3 & 840 &  17 & 1.5 & 135 & 2.6 \\
 \hline \hline
\end{tabular}
\end{table*}

\begin{figure*}
\centering
\begin{tabular}{c@{\hspace{1cm}}c}
\includegraphics[width=6.0cm]{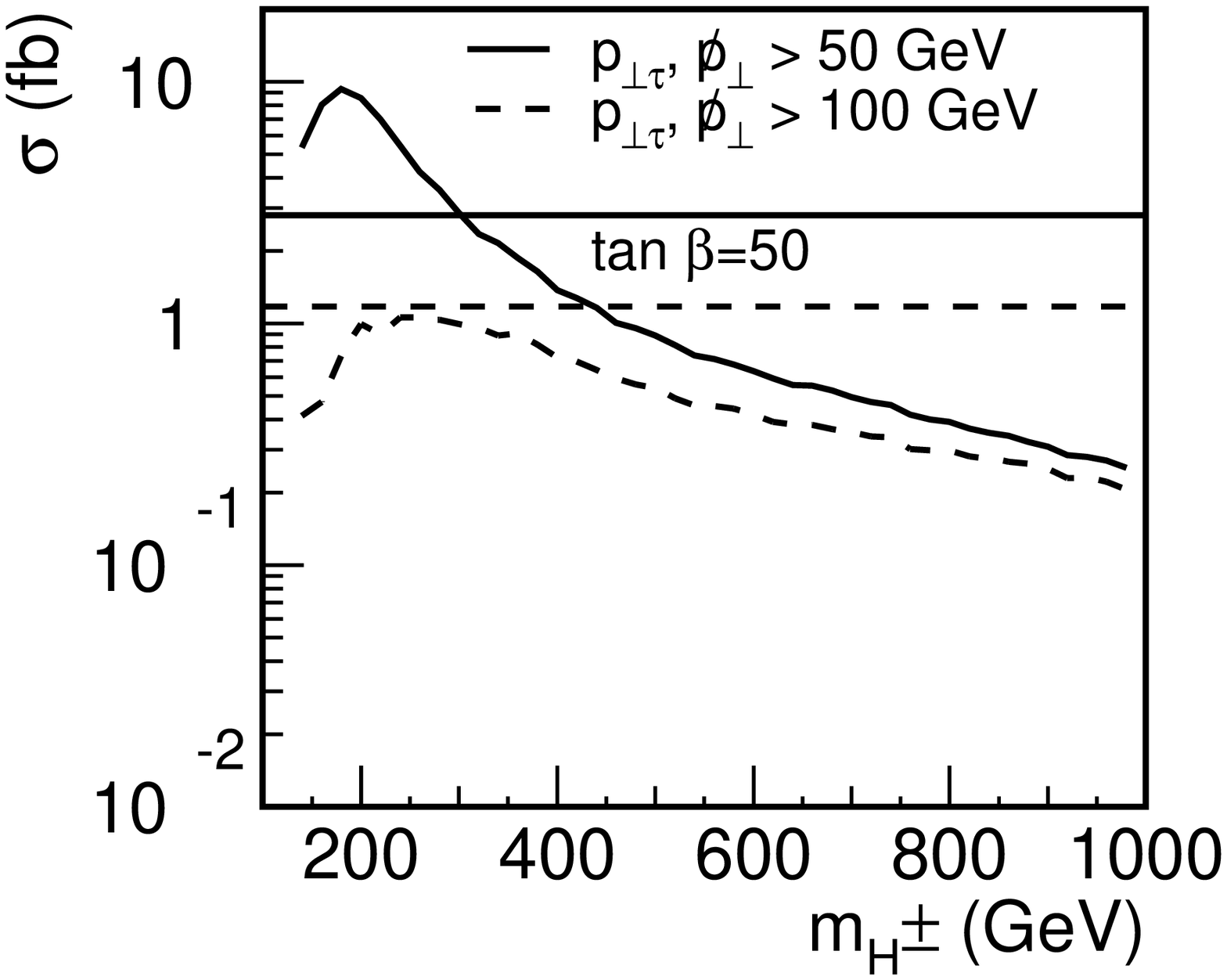} &
\includegraphics[width=6.0cm]{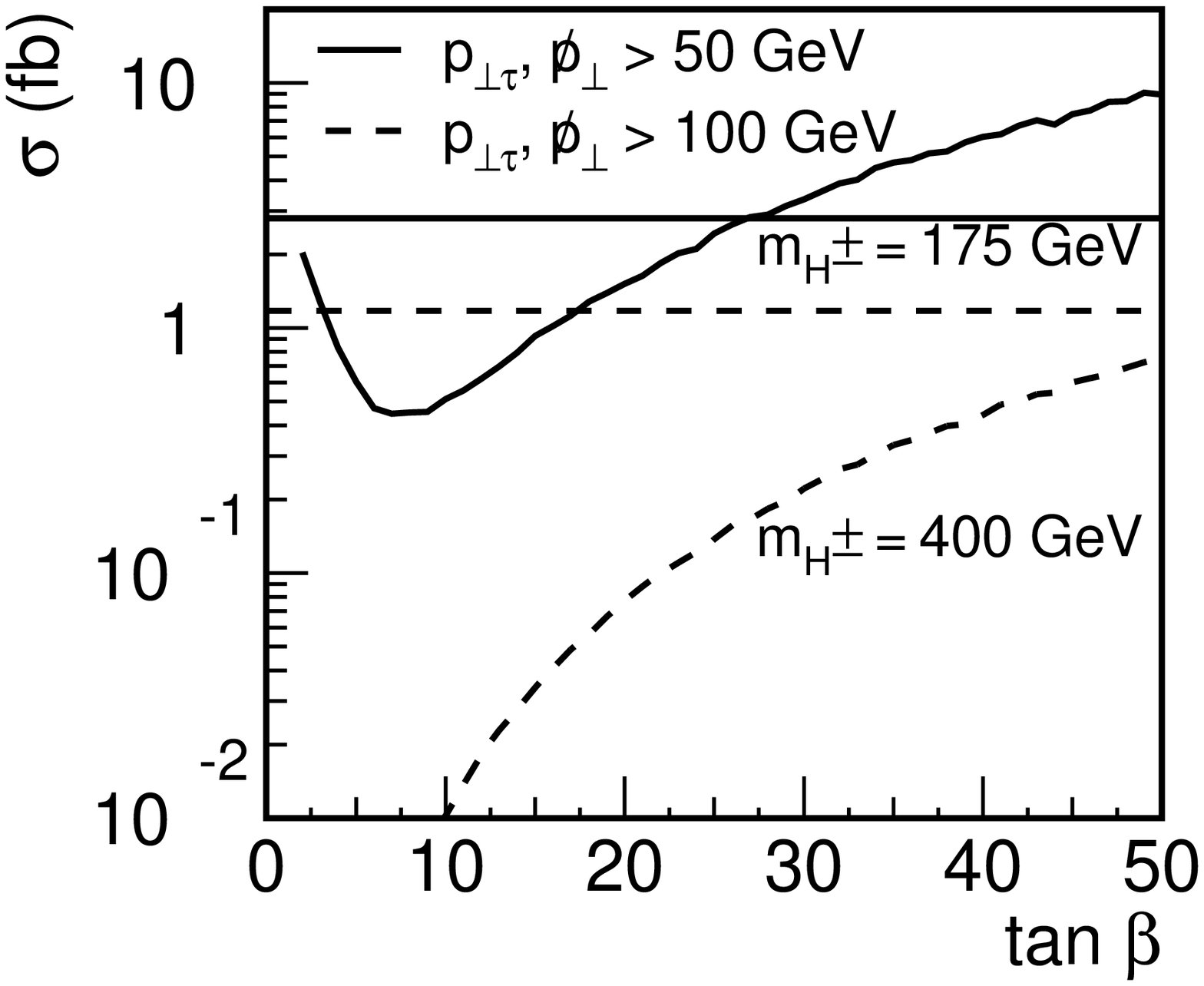}\\[4mm]
\includegraphics[width=6.0cm]{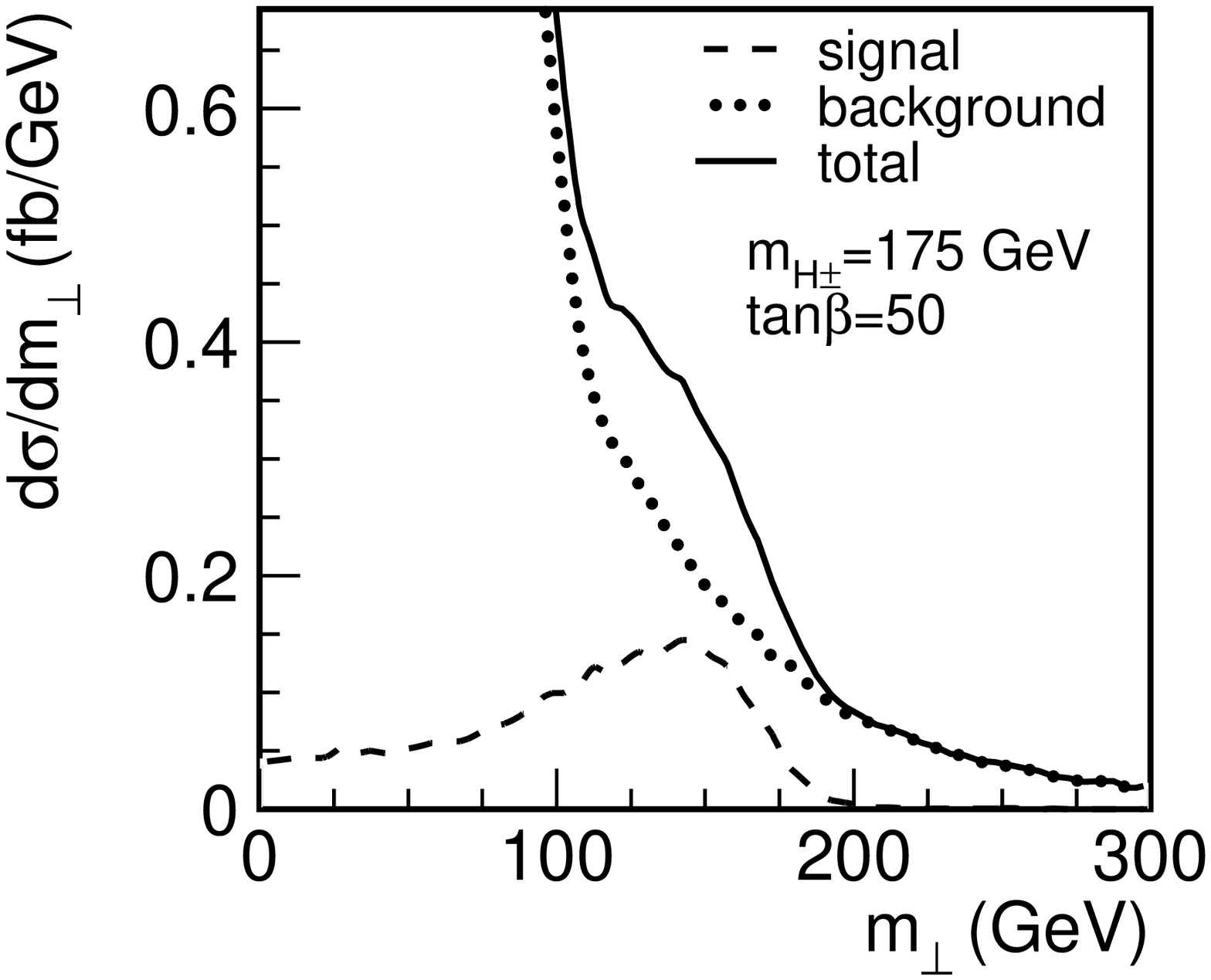} &
\includegraphics[width=6.0cm]{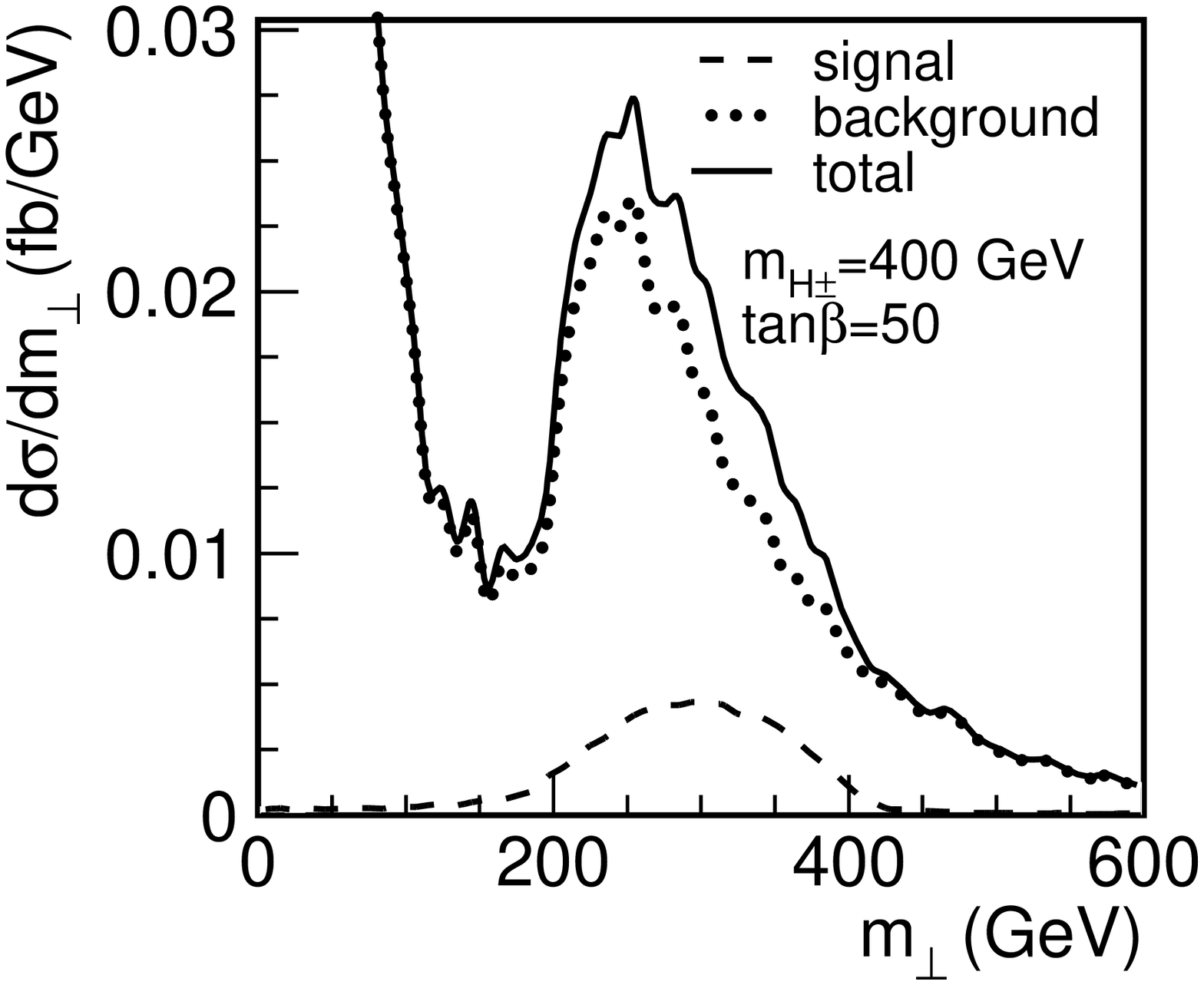}
\end{tabular}
\caption{Upper row: $H^\pm$ mass and $\tan\beta$ dependence of the integrated
  cross-section in the $m_h^\mathrm{max}$ scenario.
  Solid curves are with all cuts
  of table~\ref{cuts} and dashed curves are with the harder cuts
  $p_{\perp\tau_\mathrm{jet}},\mbox{$\!\not\!p_{\perp}$}>100$ GeV. 
  The horizontal lines correspond to $S/\sqrt{B}=5$.
  Lower row: $m_\perp$ distribution for the signal (dashed) in the
  $m_h^\mathrm{max}$ scenario with $\tan\beta=50$ and $m_{H^\pm} = 175$ GeV
  (left) as well as $m_{H^\pm} = 400$ GeV (right) together with the
  background (dotted) with all cuts of table~\ref{cuts}
  (for $m_{H^\pm} = 400$ GeV the cuts
  $p_{\perp\tau_\mathrm{jet}}, \mbox{$\!\not \!p_{\perp}$} > 100$~GeV are
  used).
}\label{mhmax}
\end{figure*}

The dominant production mechanisms for $H^\pm W^\mp$ at hadron colliders
are $b\bar{b}$ annihilation at tree-level and gluon fusion
at one-loop-level.
Here, we focus on the parameter region with intermediate
$H^\pm$ masses ($\sim m_t$)
and large $\tan\beta$, where
the decay $H^\pm \rightarrow \tau\nu_\tau$ has a large branching ratio and
where the $b\bar{b}$ annihilation dominates.
We have implemented~\cite{DEriksson} the two processes
$b\bar b \rightarrow H^+ W^-$
and $b\bar b \rightarrow H^- W^+$ as separate external processes to
{\sc Pythia}~\cite{Sjostrand:2000wi}.
For the calculation of the MSSM scenario and the corresponding Higgs
masses and the branching ratios of $H^\pm$
we use {\sc FeynHiggs} 2.2.10 \cite{Hahn:2005cu}.
We only consider hadronic $\tau$ decays,
$\tau \to \nu_\tau +  \tau_\mathrm{jet}$, which are performed using
the program {\sc Tauola}~\cite{Golonka:2003xt} in order to
properly take into account the spin effects, resulting in the
signature $2 j + \tau_\mathrm{jet} + \mbox{$\not \!p_\perp$} \,$.
The dominant irreducible SM background arises from $W$ + 2 jets
production which we have simulated with help of the package
ALPGEN \cite{Mangano:2002ea} again complemented with {\sc Tauola}
to perform the $\tau$ decay.

Since this study is performed at parton level, without any parton showering or
hadronisation, the momenta of the jets are smeared as a first
approximation to take these, as well as detector effects, into account.
After smearing the following basic cuts are applied:
$|\eta_{\tau_\mathrm{jet}}|<2.5$, $|\eta_j|<2.5$, $\Delta R_{jj}>0.4$,
$\Delta R_{\tau_\mathrm{jet} j}>0.5$, and $p_{\perp \mathrm{jet}}>20$ GeV.
We then apply the further cuts
given in table~\ref{cuts} in order to suppress the background,
where $m_\perp=
\sqrt{2p_{\perp\tau_\mathrm{jet}} \mbox{$\not \!p_\perp$}
[1-\cos(\Delta\phi)]}$
(with  $\Delta\phi$ being the azimuthal angle between
$p_{\perp\tau_\mathrm{jet}}$ and \mbox{$\not \!p_\perp$}) is the
transverse mass and $p_{\perp hj}$ ($p_{\perp sj}$) is the harder
(softer) of the two jets.
The cuts \mbox{$\not \!p_\perp, p_{\perp\tau_\mathrm{jet}} >50 $ GeV}
are primarily included to take the reducible QCD background into
account, which has not been simulated explicitly. 
In order to estimate the sensitivity of the final results
due to this choice we have also used an alternative set of harder
cuts, \mbox{$\not \!p_\perp, p_{\perp\tau_\mathrm{jet}} >100 $ GeV}.

\section{Standard MSSM scenarios}

The signal cross-sections in the standard $m_h^\mathrm{max}$ scenario
as well as the resulting number of events and the significance
$S/\sqrt{B}$ are given in table~\ref{cuts} for the MSSM
parameters 
$\mu=200$ GeV, $M_\mathrm{SUSY}=1$ TeV, $A_t=A_b=A_\tau=2$ TeV,
$M_2=200$ GeV, and $m_{\tilde g}=800$ GeV.
The upper row in fig.~\ref{mhmax} shows the
$m_{H^\pm}$ and $\tan\beta$ dependence of the cross-section after all cuts
of table~\ref{cuts} (solid curves) and after the harder cuts
$p_{\perp\tau_\mathrm{jet}}, \mbox{$\!\not \!p_{\perp}$}>100$ GeV
(dashed curves).
The horizontal lines indicate the cross-section needed for
$S/\sqrt{B}=5$, corresponding to
 $\tan\beta \gtrsim 30$ if $m_{H^\pm}=175$~GeV and
$150~\textrm{GeV} \lesssim m_{H^\pm} \lesssim 300 $~GeV  if $\tan\beta =50$
with the softer cuts
$p_{\perp\tau_\mathrm{jet}}, \mbox{$\!\not\!p_{\perp}$}>50 $ GeV,
whereas with the harder cuts
$\tan\beta$ has to be larger than at least 50.

The lower row in fig.~\ref{mhmax} shows the resulting
$m_\perp$ distribution for $m_{H^\pm} = 175$~GeV as well as
$m_{H^\pm} = 400$~GeV in the case $\tan\beta =50$ compared to the
background after
all cuts in table~\ref{cuts} have been applied.
In the high mass case the harder cuts
$p_{\perp\tau_\mathrm{jet}}, \mbox{$\!\not \!p_{\perp}$} > 100$~GeV
are used giving $S/\sqrt{B}=3.2$.
Applying an upper cut  $m_\perp < 200$~GeV ($m_\perp < 500$~GeV)
for $m_{H^\pm} = 175$~GeV ($m_{H^\pm} = 400$~GeV)
only marginally improves $S/\sqrt{B}$ from 17 (3.2) to 19 (3.3).
In the same figure we also see that the harder cuts create a fake peak in the
background. Finally, using the harder cuts
$p_{\perp\tau_\mathrm{jet}}, \mbox{$\!\not \!p_{\perp}$} > 100$~GeV
the significance
for $m_{H^\pm} = 175$~GeV and $\tan\beta =50$ is reduced to
 $S/\sqrt{B}=3.1$. However, in this case using an upper cut
 $m_\perp < 200$~GeV is beneficial leading to a significance of
 $S/\sqrt{B}=6.4$.

The above results in standard MSSM scenarios depend only weakly on the
underlying MSSM parameter point except for parameter points with large
mass splitting between charged and neutral Higgs bosons which will be
described in the next section.

\section{Resonant scenarios}

In MSSM parameter regions with $|\mu|, |A_t|,$ or $|A_b| > 4 M_\mathrm{SUSY}$
the dominant terms of the 1-loop corrections to the quartic couplings
in the Higgs sector \cite{Pilaftsis:1999qt}
can induce a large mass splitting
between the charged and neutral Higgs bosons
\cite{Akeroyd:2001in,mohngollubassamagan:2005}.
A scan over the relevant parameters shows that scenarios with
$m_A > m_{H^\pm} + m_W$ are possible in  a large range in both
$\tan\beta$ and $m_{H^\pm}$ \cite{Eriksson:2006yt}.
On average $0.2\%$ of the scanned MSSM scenarios as outlined in
table \ref{resonantpar} turn out to fulfil
this relation.
In this case the cross section for $b\bar b \rightarrow H^\pm W^\mp$ can be
considerably enhanced through the resonance of a Higgs
boson in the $s$-channel.

\begin{table}
\caption{MSSM parameters for the resonant scenario in addition to 
 $M_L^3 = M_E^3 = 500$ GeV, $A_t = A_b = 0$, $M_2 = m_{\tilde g} = 500$ GeV 
as well as the 
range for the scan of parameters together with the step size.}\label{resonantpar}
\centering
\begin{tabular}{lcccccc} \hline \hline
 &\multicolumn{6}{c}{MSSM parameters. All masses in GeV.}\\ 
  &
$m_{H^\pm}$ &
$\tan\beta$ &
$\mu$ & 
$M_Q^3$ &
$M_U^3$ &
$M_D^3$ 
  \\ \hline
Res. scen. &
175 
& 11 
& 3300 
 & 250 
 & 250 
 & 400  \\ 
Scan $\min$ &
100 &
1 &
1800 &
150 &
150 &
150  \\
Scan $\max$ &
450 &
40 &
3300 &
650 &
650 &
650 \\
Step size &
25 &
1 &
250  &
50  &
50  &
50  
\\
\hline \hline
\end{tabular}
\end{table}

\begin{figure}
\centering
\includegraphics[width=6.0cm]{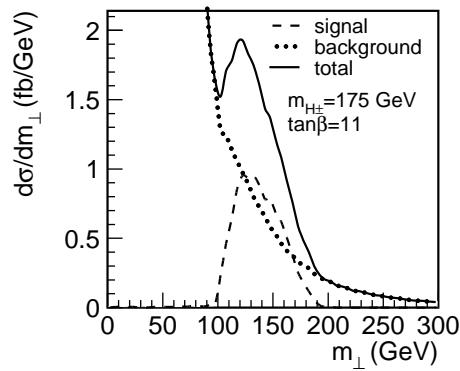}
\caption{The $m_\perp$ distribution of the signal (dashed)
  in the resonant scenario of table~\ref{resonantpar}
  and of the background (dotted)
  with all cuts of table~\ref{cuts} except
  $p_{\perp hj} > 50$~GeV and $p_{\perp sj} > 25$~GeV.}\label{resonant}
\end{figure}

Figure~\ref{resonant} shows the resulting $m_\perp$-distribution
compared to the background in the resonant scenario of table
\ref{resonantpar}. 
In case of resonant production, the $H^\pm$ and $W$ bosons are produced with
typically small transverse momenta. Thus it is favourable to loosen the cuts
on the light jets from the $W$.
Applying the basic and additional cuts from
table~\ref{cuts},
except the cuts $p_{\perp hj} > 50$~GeV and $p_{\perp sj} > 25$~GeV
on the light jets, we get an integrated cross-section of 52~fb
for a charged Higgs boson mass of 175~GeV corresponding to
a significance $S/\sqrt{B}=56$.
For comparison, if we apply the harder cuts
$p_{\perp\tau_\mathrm{jet}} > 100$~GeV and
 $\mbox{$\!\not \!p_{\perp}$}>100$~GeV
in this resonant case the significance is reduced drastically
to $S/\sqrt{B}=0.2$ due to the typically
small transverse momentum of the $H^\pm$-boson. Thus, in the case of harder
cuts the resonantly enhanced cross-section is only of use if $m_{H^\pm}$
is large enough such that $m_{H^\pm}/2$ is well above $100$ GeV.

The study of these resonant scenarios
also illustrates what could happen in a
general 2HDM, where the Higgs boson masses are more or less
independent parameters and the resonant scenarios are less fine-tuned.

\section{MSSM with complex parameters}

\begin{figure*}
\centering
\begin{tabular}{c@{\hspace{1cm}}c}
\includegraphics[width=6.0cm]{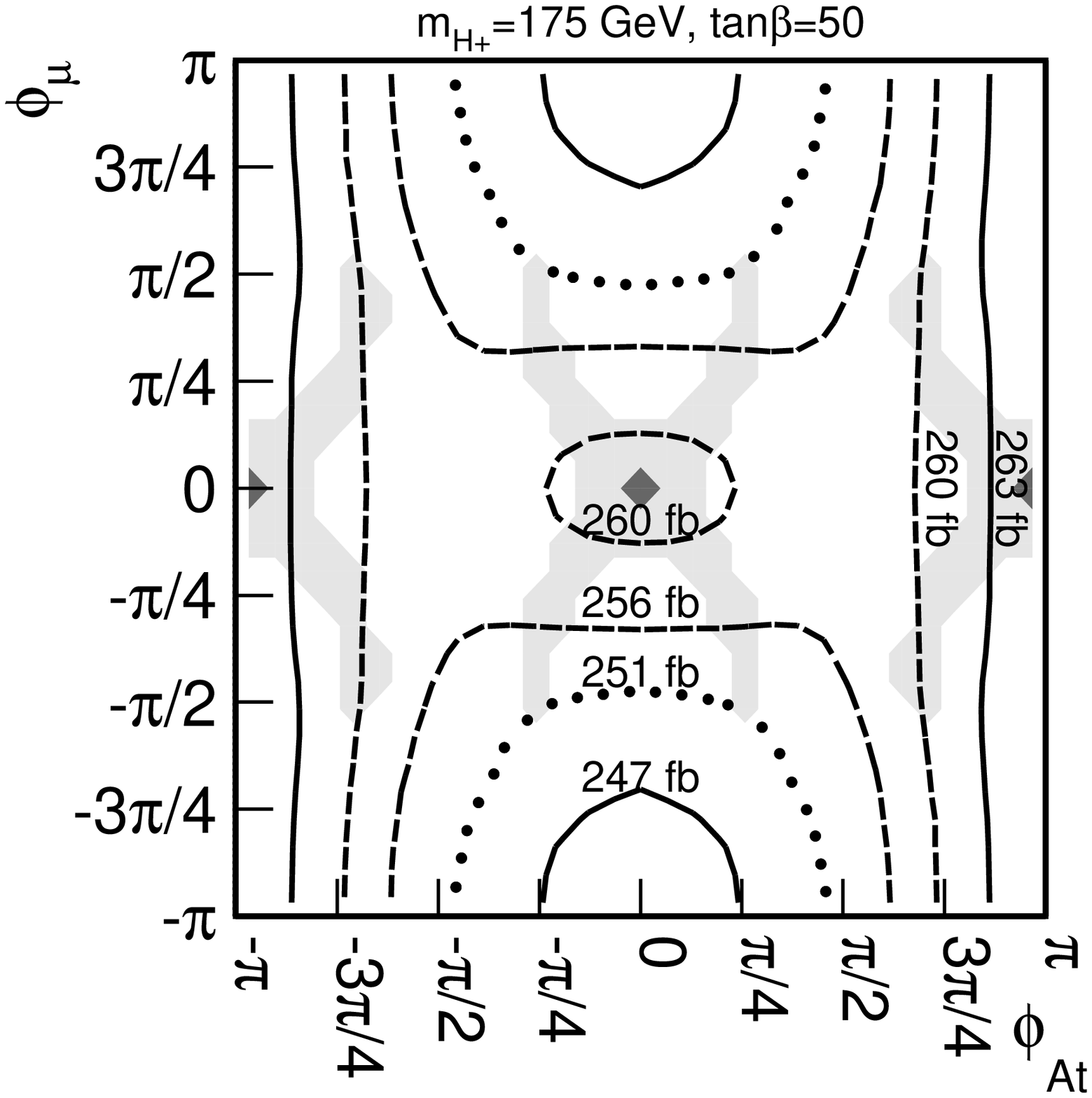} &
\includegraphics[width=6.0cm]{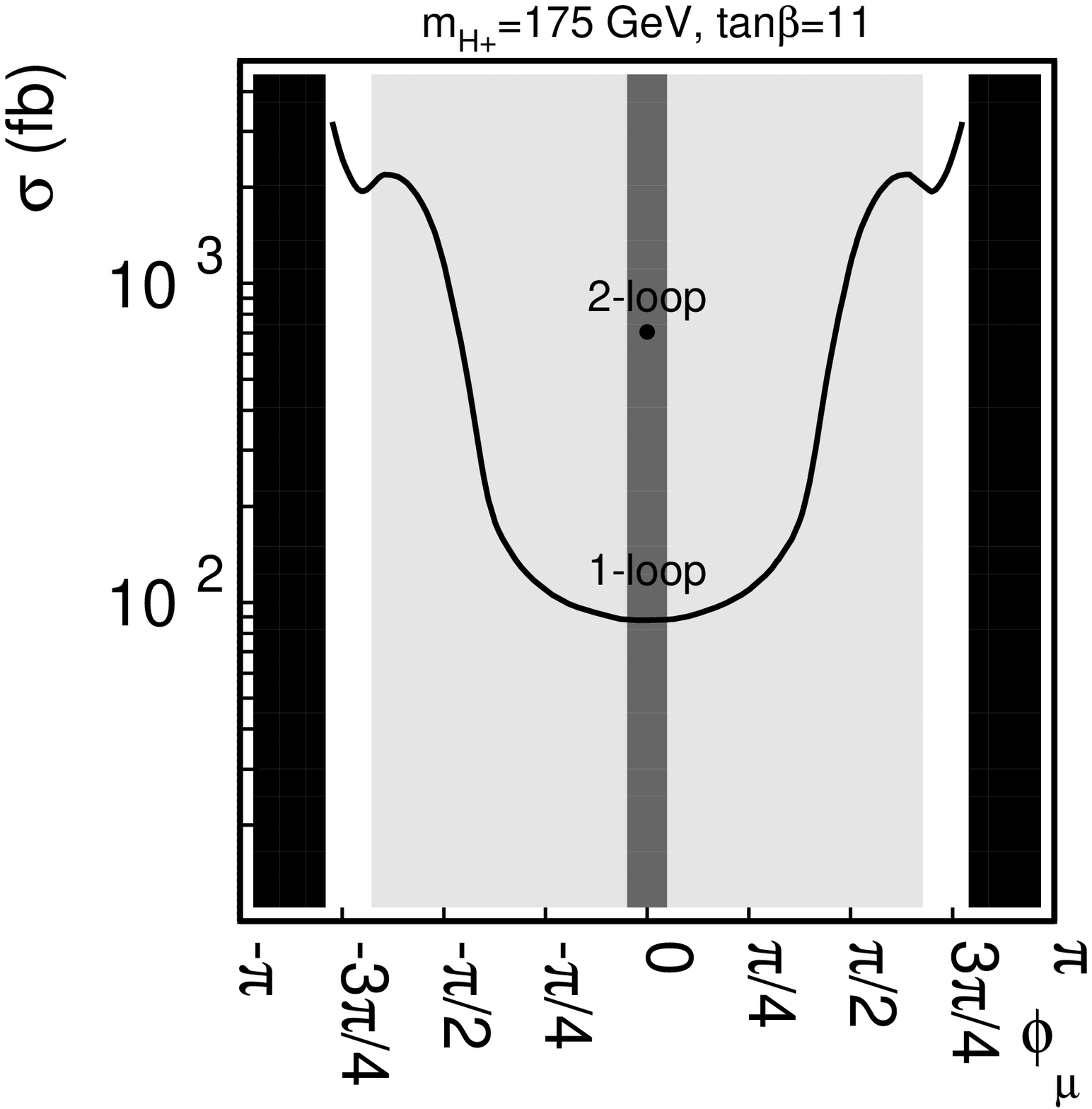}
\end{tabular}
\caption{Left plot: Contours of the total cross-section in
  the $m_h^\mathrm{max}$ scenario.
  Right plot: Total cross-section in the resonant scenario of
  table~\ref{resonantpar} with the Higgs masses etc.\ calculated to
  one-loop. For comparison the result with all available
  corrections for $\phi_\mu=0$ is also shown.
  The light shaded areas are in
  agreement with the constraints from $a_\mu$ and $\delta\rho_0$ and
  the dark shaded ones
  are also in agreement with the constraints from EDMs.
  In the black areas numerical instabilities occurred in the calculation of
  the Higgs masses and mixing matrix.}\label{cpv}
\end{figure*}

In the general MSSM many parameters can be complex.
Then explicit CP violation occurs also in the Higgs sector through
loops of supersymmetric particles~\cite{Pilaftsis:1999qt,Pilaftsis:1998dd},
which affects our signal process because
of the neutral Higgs bosons exchanged in the $s$-channel.
The leading contributions to CP violation in the neutral Higgs
sector arise from loops of the scalar top and (to a lesser extend)
of the scalar bottom sector
where the possibly complex Higgs/higgsino mass parameter $\mu$
and the trilinear scalar couplings  $A_t$ and $A_b$ are dominant.
Hence, assuming $\phi_{A_b} = \phi_{A_t}$
we concentrate in the following on
the phases $\phi_\mu$ and $\phi_{A_t}$ of $\mu$ and $A_t$, respectively, which
have the largest effect on the neutral Higgs
sector and thus possibly affect our signal.
Despite the fact that the SUSY phases may be severely constrained by
bounds on the Electric Dipole Moments (EDMs), these constraints are rather
model dependent and may be evaded in scenarios with heavy first and second
generation sfermions, due to cancellations among various contributions
to the EDMs or due to additional contributions from lepton flavour
violating terms in the Lagrangian, for a review see e.g.~\cite{Olive:2005ru}.
Thus, we have varied $\phi_\mu$ and
$\phi_{A_t}$ independently in the range $-\pi < \phi < \pi$ in order
to investigate the phase dependence of our signal.

In standard MSSM scenarios, such as the $m_h^\mathrm{max}$ scenario,
we find only small ($\sim 5 \%$)
$\phi_\mu$ and $\phi_{A_t}$ dependencies of our signal cross-section
as shown in the left plot in fig.~\ref{cpv}.
The right plot in fig.~\ref{cpv}
shows the dependence of the cross-section on
$\phi_\mu$ in the resonant scenario of table \ref{resonantpar},
where the Higgs masses, couplings and widths have been
calculated with {\sc FeynHiggs}  at one-loop accuracy\footnote{
A calculation with all available corrections is not possible here because the
phases lead to numerical instabilities.}.
The very large phase dependence in this scenario
is due to the fact that the production goes
from non-resonant to resonant when the phase is varied.
More specifically,
we get $m_A = m_{H_3} = 246$~GeV for $\phi_\mu = 0$ in the 1-loop case, which
is below the resonant
threshold, whereas $m_{H_3}=342$ GeV for the largest values of $\phi_\mu$
where we got a stable result, which is clearly in the
resonant regime.

In principle, the process under study also offers the possibility to explore
effects of CP violation via a
CP-odd rate asymmetry $A_\mathrm{CP}$
between $\sigma(pp \rightarrow H^+ W^-)$ and 
$\sigma(pp \rightarrow H^- W^+)$.
However, in all considered MSSM scenarios we find that 
$A_\mathrm{CP}$ is smaller than about $1\%$.

\begin{acknowledgement}
This work has been supported by the G\"oran Gustafsson Foundation.
\end{acknowledgement}


\begin{thebibliography}{999}

\bibitem{Barnett:1987jw}
R.~M. Barnett, H.~E. Haber, and D.~E. Soper,
  Nucl. Phys. {\bf B306} (1988) 697.

\bibitem{Bawa:1989pc}
A.~C. Bawa, C.~S. Kim, and A.~D. Martin,
  Z. Phys. {\bf C47} (1990) 75.

\bibitem{Borzumati:1999th}
F.~Borzumati, J.-L. Kneur, and N.~Polonsky,
  Phys. Rev. {\bf D60} (1999) 115011, hep-ph/9905443.

\bibitem{Miller:1999bm}
D.~J. Miller, S.~Moretti, D.~P. Roy, and W.~J. Stirling,
  Phys. Rev. {\bf D61}, (2000) 055011, hep-ph/9906230.

\bibitem{Alwall:2004xw}
J.~Alwall and J.~Rathsman,
  JHEP {\bf 0412} (2004) 050, hep-ph/0409094.

\bibitem{Mohn:2007fd}
B.~Mohn, M.~Flechl, and J.~Alwall,
  ATLAS discovery potential for the Charged Higgs
     Boson in $H^{+} \to \tau \nu$ decays,
  ATL-PHYS-PUB-2007-006, 2007.

\bibitem{Hesselbach:2007jj}
  S.~Hesselbach, S.~Moretti, J.~Rathsman and A.~Sopczak,
  arXiv:0708.4394 [hep-ph].

\bibitem{Dicus:1989vf}
D.~A. Dicus, J.~L. Hewett, C.~Kao, and T.~G. Rizzo,
  Phys. Rev. {\bf D40} (1989) 787.

\bibitem{BarrientosBendezu:1998gd}
A.~A. Barrientos~Bendezu and B.~A. Kniehl,
  Phys. Rev. {\bf D59} (1999) 015009, hep-ph/9807480.

\bibitem{Moretti:1998xq}
S.~Moretti and K.~Odagiri,
  Phys. Rev. {\bf D59} (1999) 055008, hep-ph/9809244.

\bibitem{Brein:2000cv}
O.~Brein, W.~Hollik, and S.~Kanemura,
  Phys. Rev. {\bf D63} (2001) 095001, hep-ph/0008308.

\bibitem{Asakawa:2005nx}
E.~Asakawa, O.~Brein, and S.~Kanemura,
  Phys. Rev. {\bf D72} (2005) 055017, hep-ph/0506249.

\bibitem{Eriksson:2006yt}
D.~Eriksson, S.~Hesselbach and J.~Rathsman,
  accepted by Eur.\ Phys.\ J. {\bf C},
  hep-ph/0612198.

\bibitem{Gao:2007wz}
J.~Gao, C.~S.~Li and Z.~Li,
  arXiv:0710.0826 [hep-ph].

\bibitem{DEriksson}
D.~Eriksson,
  The fortran code for the $b\bar{b}\to H^+W^-$ and $b\bar{b}\to
  H^-W^+$ processes are available from
  \verb+http://www.isv.uu.se/thep/MC/pybbwh/+, including a manual.

\bibitem{Sjostrand:2000wi}
T.~Sj\"ostrand {\em et~al.},
  Comput. Phys. Commun. {\bf 135} (2001) 238, hep-ph/0010017.

\bibitem{Hahn:2005cu}
T.~Hahn, W.~Hollik, S.~Heinemeyer, and G.~Weiglein,
  hep-ph/0507009.

\bibitem{Golonka:2003xt}
P.~Golonka {\em et~al.},
  hep-ph/0312240.

\bibitem{Mangano:2002ea}
M.~L. Mangano, M.~Moretti, F.~Piccinini, R.~Pittau, and A.~D. Polosa,
  JHEP {\bf 0307} (2003) 001, hep-ph/0206293.

\bibitem{Pilaftsis:1999qt}
A.~Pilaftsis and C.~E.~M. Wagner,
  Nucl. Phys. {\bf B553} (1999) 3, hep-ph/9902371.

\bibitem{Akeroyd:2001in}
A.~G. Akeroyd and S.~Baek,
  Phys. Lett. {\bf B525} (2002) 315, hep-ph/0105228.

\bibitem{mohngollubassamagan:2005}
B.~Mohn, N.~Gollub, and K.~A. Assamagan,
  Study of the $H^\pm \to W^\pm H^0$ decay in a large mass splitting MSSM
         scenario with ATLAS,
  ATL-PHYS-PUB-2005-017.

\bibitem{Pilaftsis:1998dd}
A.~Pilaftsis,
  Phys. Lett. {\bf B435} (1998) 88, hep-ph/9805373.

\bibitem{Olive:2005ru}
  K.~A.~Olive, M.~Pospelov, A.~Ritz and Y.~Santoso,
  Phys.\ Rev. {\bf D72} (2005) 075001,
  hep-ph/0506106.

\end{thebibliography}
\end{document}